\documentclass[9pt,twocolumn,twoside]{osajnl}

\journal{ol} % Choose journal (ao, aop, josaa, josab, ol)

\setboolean{shortarticle}{true} 
\ifthenelse{\boolean{shortarticle}}{\colorlet{color2}{color2b}}{\colorlet{color2}{color2}} % Automatically switches colors for short articles

\title{Phase diffusion quantum entropy source on a silicon chip}

\author[1,*]{Miquel Rud\'e}
\author[1]{Carlos Abell\'an}
\author[1]{Albert Capdevila}
\author[2]{David Domenech}
\author[1,3]{Morgan W. Mitchell}
\author[1]{Waldimar Amaya}
\author[1,3]{Valerio Pruneri}

\affil[1]{ICFO-Institut de Ciencies Fotoniques, The Barcelona Insitute of Science and Technology, 08860 Castelldefels, Barcelona,Spain}
\affil[2]{VLC Photonics S.L Cami de Vera s/n, Edificio 9B, Valencia, Spain}
\affil[3]{ICREA-Institucio Catalana de Recerca i Estudis Avançats, 08010 Barcelona, Spain}

\affil[*]{Corresponding author: miquel.rude@icfo.eu}

\dates{Compiled \today}

\ociscodes{(130.0130) Integrated optics; (230.3120) Integrated optics devices; (270.0270) Quantum optics.}

\doi{\url{http://dx.doi.org/10.1364/optica.XX.XXXXXX}}

\begin{abstract}

We report an accelerated laser phase diffusion quantum entropy source with all non-laser optical and optoelectronic elements implemented in silicon photonics. The device uses efficient and robust single-laser accelerated phase diffusion methods, and implements the whole quantum entropy source scheme including an unbalanced Mach-Zehnder interferometer with optimized splitting ratio, in a 0.5 mm$\times$1 mm footprint. We demonstrate Gbps raw entropy-generation rates in a technology compatible with conventional CMOS fabrication techniques.

\end{abstract}

\setboolean{displaycopyright}{true}

\begin{document}

\maketitle
\thispagestyle{fancy}
\ifthenelse{\boolean{shortarticle}}{\abscontent}{}

\section{Introduction}

On-demand generation of random numbers (RNs) is a key ingredient for fields as diverse as Monte Carlo simulations \cite{Cai2007, Click2011}, online gambling applications \cite{Hall1997}, decision making algorithms, cybersecurity \cite{Shannon1949, Tajima2007}, and even tests of fundamental physics \cite{Abellan2015, Hensen2015, Giustina2015, Shalm2015}. Although pseudo-RNs can be easily generated using computational algorithms, true RNs can only be created using physical processes \cite{VonNeumann1951, Tajima2007}. Quantum entropy sources (QESs) make use of the intrinsic randomness of quantum mechanics to create strings of random bits. Several implementations of QESs have been demonstrated, including splitting of single photons \cite{Rarity1994, Jennewein2000}, photon arrival time \cite{Wahl2011}, vacuum fluctuations \cite{Gabriel2010}, laser chaos \cite{Argyris2010, Ugajin2017} , and phase diffusion (PD) in laser diodes \cite{Qi2010, Jofre2011, Abellan2014, Yuan2014, Nie2015}. In particular, PD-QESs have been shown to achieve high bit rates and offer strong randomness guarantees \cite{Abellan2015}. 

For future devices it is desirable to scale down these bulky technologies into integrated devices. Recently, an integrated QES using a light emitting diode (LED) and a single-photon avalanche photodetector (SPAD) achieved 1 Mbps bit rates \cite{Khanmohammadi2015}. PD-QESs have the potential to achieve several orders of magnitude higher rates as they use conventional photodetectors instead of SPAD. A PD-QES in an indium phosphide (InP) integrated circuit was demonstrated with Gbps rates \cite{Abellan2016}. Implementation in silicon photonics, which we show here, allows direct integration with conventional complementary metal oxide semiconductor (CMOS) electronics, enabling QES deployment in the most advanced semiconductor industry.

In this work we demonstrate a PD-QES on a Si chip using an integrated unbalanced Mach-Zehnder interferometer (uMZI) scheme. The laser component cannot be directly implemented in Si photonics, although hybrid technologies, like Germanium-on-Silicon \cite{Camacho2012} or 2D materials \cite{Bie2017} have shown the potential for full PD-QES integration onto a single chip. Here the device is driven by  an external DFB laser operated in gain-switching (GS) mode to generate pulses with equal amplitudes and random initial phases. These pulses are then interfered in the uMZI, thus creating a train of pulses with random amplitudes that are measured in a high bandwidth integrated photodetector (PD). The scheme  shows high stability over time and can potentially deliver Gbps bit rates with appropiate digitization components. 

\section{Experiment}

\begin{figure}[htbp!]
\centering
\includegraphics[width=\linewidth]{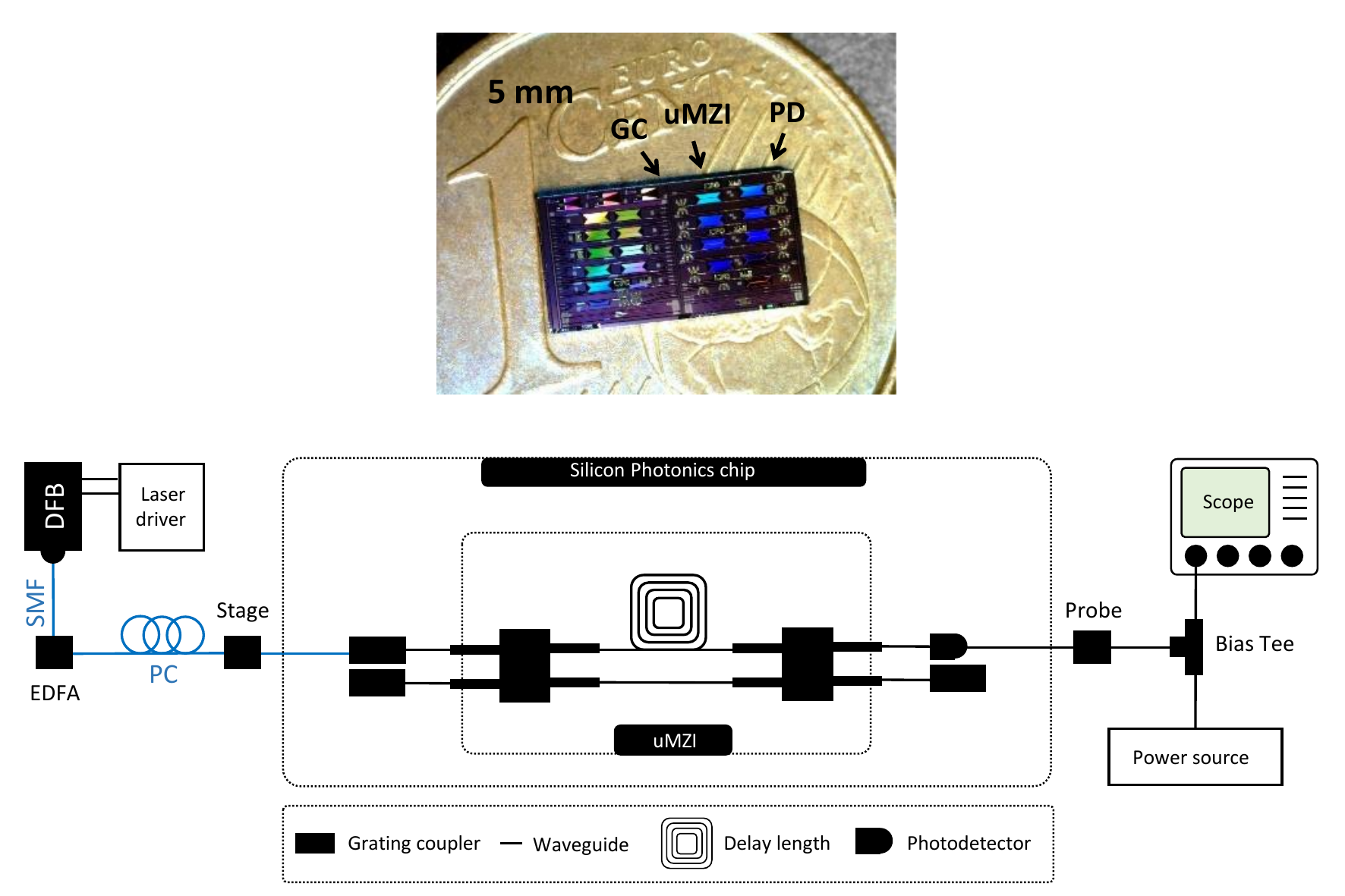}
\caption{Upper: Microscope image of the 4 mm x7.5 mm Si photonic chip with a one euro cent coin for reference. One such chip can contain up to 20 PD-QESs. A single PD-QES block is less than 0.5 mmx1 mm. Lower: Schematic of the experimental set-up. The DFB laser is biased using a current source and modulated with a 1 GHz sinusoidal wave. The output pulses with random phases are amplified using an erbium-doped fiber amplifier (EDFA) and coupled into a single mode Si waveguide using an SMF and a GC. The pulses are split in an MMI, with a power splitting ratio of 0.02:0.98, to the two arms of an uMZI, in order to compensate the losses introduced by the longer path. The interference signal is detected using an integrated photodetector and analyzed with a real-time oscilloscope.}
\label{Fig1}
\end{figure}

Figure~\ref{Fig1} shows a schematic of the experimental set-up and an image of the Si chip containing QES devices. The Si chip implements both the interferometry and photodetection elements of the PD-QES strategy. The laser component is interfaced to the chip by a grating coupler (GC). A single-frequency ($\lambda$ = 1550 nm) DFB laser is operated in GS mode, with a mean drive current of 14 mA and a sinusoidal modulation at 1 GHz, applied via a bias-tee.  As the laser threshold is 10 mA, this takes the laser above and far below threshold on each cycle, producing a train of linearly-polarized optical pulses of duration $\sim$300 ps. 

Due to phase diffusion, subsequent pulses have random relative phases, while also having the same waveforms. In order to couple the pulses into the Si chip, the laser output is directed via an SMF towards a GC at 10$^\circ$ incidence using a 6-axis micropositioner. A polarization controller (PC) is used to adjust the input polarization to minimize the coupling losses due to the GC, which were estimated to be $\sim$7 dB by measuring the transmission through a straight waveguide. 

In the chip, a first multimode interference coupler (MMI) splits the input light, with a power-splitting ratio of 0.02:0.98, to the two arms of an uMZI, in order to compensate the losses introduced by the longer path. The stronger output experiences the longer path, and the two arms are re-combined at a second MMI, with splitting ratio 50:50. The splitting ratio ($t_l/t_s$) of the first MMI , where $t_l$ ($t_s$) is the transmission to the long (short) arm of the uMZI, is given by 
\begin{equation}
t_l/t_s=\text{exp}(\kappa\Delta l)
\label{Eq2}
\end{equation}
here $\kappa\sim$ 0.56 cm$^{-1}$ is the attenuation coefficient of the Si waveguide and $\Delta l$ = 6.9 cm is the relative path difference of the uMZI, given by $\Delta l = \tau c/{n_{g}}$, where $\tau$ = 1 ns is the pulse repetition rate and $n_{g}\sim$ 4.3 is the effective group refractive index of the Si waveguide. This implies a relative attenuation by a factor of $\sim$50, compensated by the first MMI and thus equalizing the field strength reaching the detector, while the path length difference introduces a delay of 1 ns that creates the conditions for temporal overlap of subsequent pulses. Careful control of these parameters is crucial in order to obtain high interference visibility. Finally, the interfered pulses are detected by a fast (10 GHz) on-chip photodiode (responsivity $\sim$ 0.7 A/W) and sent to a 4 GHz real-time oscilloscope via a bias-tee. 

\section{Discussion}

As described in \cite{Abellan2015}, the power detected by the integrated photodiode is given by
\begin{equation}
P_{\rm det}(t) = P_l(t) + P_l(t+\tau) + 2{\cal V}\sqrt{P_l(t) P_l(t+\tau) }\cos({\Delta\theta} + {\Delta\phi})
\label{Eq1}
\end{equation}
where $P_l(t)$ is the instantaneous laser power at time $t$, $\tau =$ 1 ns is the pulse repetition period, ${\cal V}$ is the interference visibility, $\Delta \theta$ is the relative phase between subsequent pulses, and $\Delta \phi$ is the optical phase acquired in the uMZI.  Due to strong phase diffusion in the time below threshold, the statistical description of $\Delta \theta$ is, to a very good approximation, random, i.e., uniformly distributed on $[0,2 \pi)$. As a result, the cosine of $\Delta \theta$ follows a bimodal distribution \cite{Abellan2015}, irrespective of $\Delta \phi$. 

\begin{figure}[htbp!]
\centering
\includegraphics[width=\linewidth]{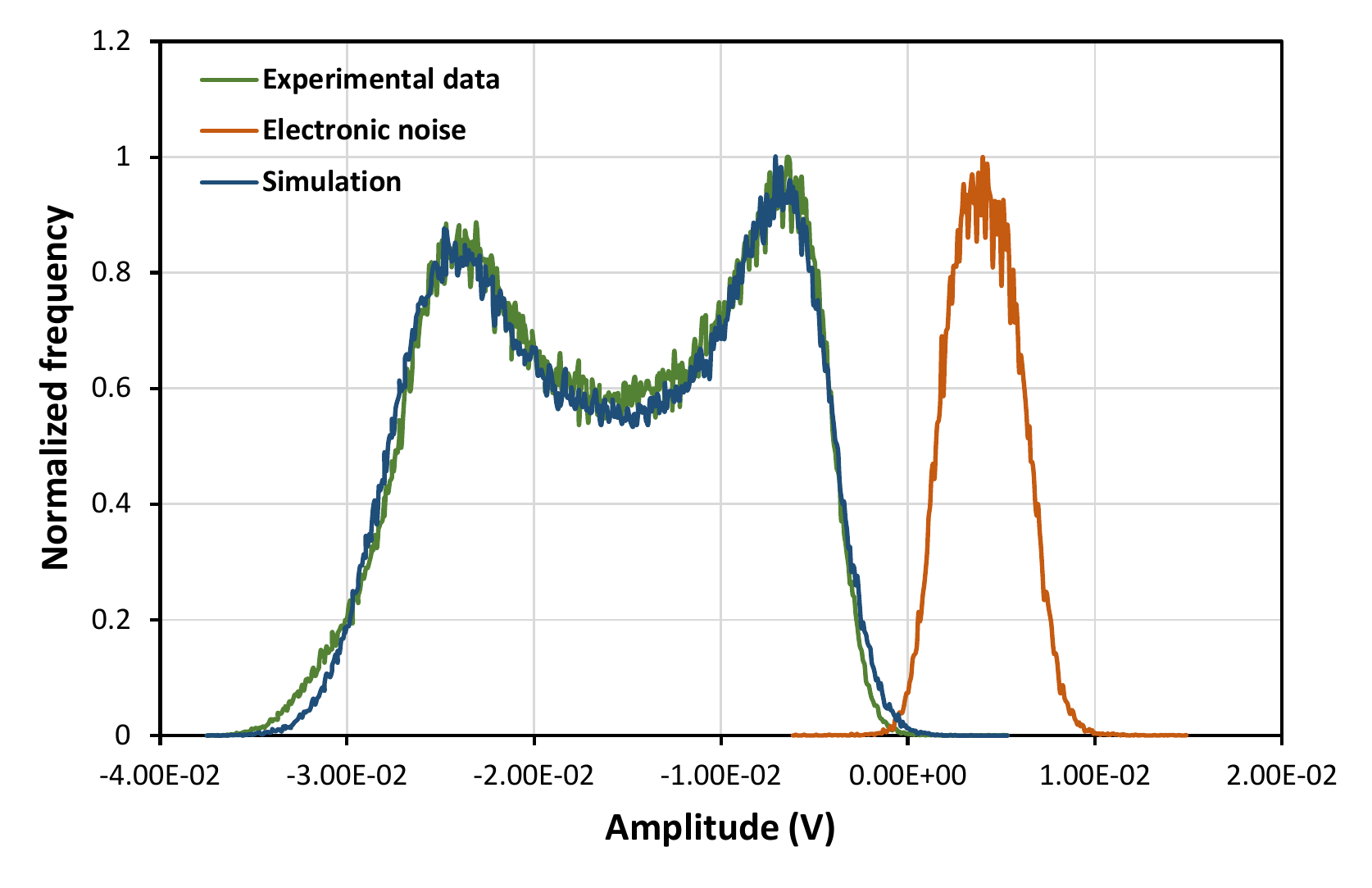}
\caption{Histogram of the electrically measured interference signal (green) and noise (orange), and simulation of the expected distribution (blue). Due to the randomness of the initial phase the accumulated signal follows a bimodal probability distribution. The histograms are obtained by continuously sampling the signal at a fixed delay point over a period of a few hours, which corresponds to approximately 7 $\times$ 10$^5$ samples. As confirmed by simulation (black curve), the bimodal distribution corresponds to Eq. \ref{Eq2} with ${\cal V} =0.74$ and random $\Delta \theta$.}
\label{Fig2}
\end{figure}

By measuring the statistics of the electrical signal when the laser is below threshold (orange curve in Figure~\ref{Fig2}) one can also obtain information about the overall noise of the system, which ultimately determines the quality of the device. Figure~\ref{Fig2} shows the observed distribution of output powers. The optical and electronic noises produce a monomodal distribution for the equivalent input power, whereas the interference process produces a strongly bimodal distribution, reflecting the arcsine distribution expected from the phase diffusion process smoothed by convolution with the electronic noise distribution. 

We also run a Monte Carlo calculation to find the parameters of Eq.~\ref{Eq1} that best fit with the observed distribution. The waveguide losses were set to $\alpha_{wg}$$\sim$ 3 dB/cm, as estimated experimentally. The electronic noise is a gaussian with rms width $\sigma_{n}$ = 1.9 mV, determined from the orange curve in Figure~\ref{Fig2}. By leaving the interference visibility $\cal V$, as well as the amplitude noise acquired in each path, i.e. the standard deviation of $P_l(t) = P(t + \tau)$, as free parameters, we found that the distribution is consistent with Eq.~\ref{Eq1}, with $P_l(t) = P_l(t + \tau)$, ${\cal V} =0.74$, and random $\Delta \theta$.

\begin{figure}[htbp!]
\centering
\includegraphics{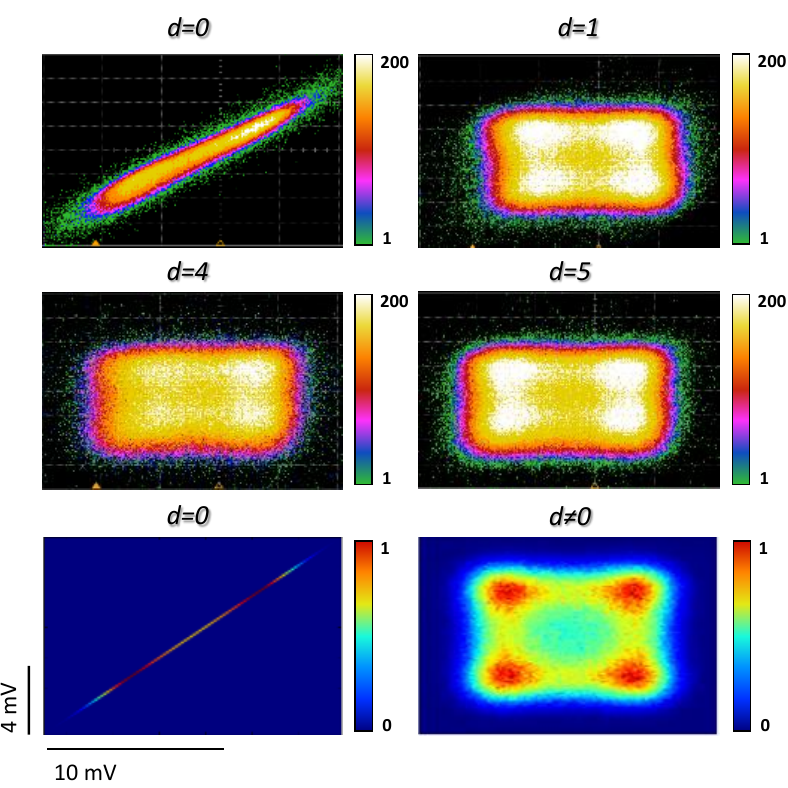}
\caption{Upper: XY plot recorded in the real-time oscilloscope, showing the accumulated occurences of an event ($V_n , V_{n-d}$) for four different shifts ($d=0, 1, 4, 5$). Lower: Simulation of the expected occurences (normalized) for the same values of d. The distribution is expected to be a straight line for $d=0$ and a 2D bimodal distribution for $d\neq 0$.}
\label{Fig3}
\end{figure}

The simulated distribution is shown in the blue curve of Figure~\ref{Fig2}. The mean square error between the observed and simulated curves is $\sim$ 10$^{-5}$ and the observed signal-to-noise ratio (SNR) is $\sim$ 4.1.

Ideally, a QES should output a string of uncorrelated and uniformly distributed random bits. However, a PD-QES provides a sequence of (in principle) uncorrelated pulses with a non-uniform (arcsine) distribution. A standard method to estimate the correlation of the bits produced by an RNG is to store a long sequence of digits $b_i$, and then calculate the unbiased estimator $\Gamma_{d}= <b_{i}b_{i-d}>-<b>^2$ in a post-processing step. Moreover in PD-QESs a randomness extractor is also applied to convert the non-uniform distribution into a uniform one, at the cost of losing a fraction of the bits \cite{Nisan1999}.
For situations where high speed digitization is not available, we introduce here a new strategy that allows to qualitatively verify this correlation using only the real time oscilloscope, without the need of any post-processing. Letting $V_n$  and $V_{n-d}$ denote the voltage amplitudes of the n-th and (n-d)-th pulses, we record in the oscilloscope the number of times an event ($V_n , V_{n-d}$) occurs for different shifts (d) in an XY plot. These results are then compared with a simulation of the expected distribution for different values of d, as shown in Figure~\ref{Fig3}. The distribution is expected to be a straight line for $d=0$ and a 2D bimodal distribution for $d\neq 0$, showing good agreement with the experimental results. The spread in the experimental distributions is due to sampling of the real-time oscilloscope. Two time windows are defined in which the values of $V_n$ and $V_{n-d}$ are sampled. Due to the non-zero width of these windows the values can take any of the values within them.    

\section{Conclusion}

In conclusion, we have demonstrated a PD-QES with Gbps bit rates and compatible with current CMOS technologies. The device is based on an external DFB laser operated in GS mode at 1 GHz coupled to a Si photonic chip that integrates the critical intererometry and detection components. Up to 20 PD-QESs can be integrated on a single chip with a footprint of only 4x7.5 mm$^2$, with a single PD-QES footprint of 0.5x1 mm$^2$. The amplitude of the interfered pulses follows a smoothed arcsine distribution, as expected for PD-QESs, with a visibility of ${\cal V} =0.74$ and random $\Delta \theta$. Also, we have introduced a method to qualitatively verify the correlation in the real-time oscilloscope without any need of offline post-processing.

Finally, the scheme could easily achieve tens of Gbps bit rates by using shorter uMZIs and thus faster modulation frequencies. This would in turn increase the SNR due to the decreased attenuation of the pulses inside the uMZI and allow integration of the PD-QES in chips with a smaller footprint.

\section{Funding Information}
Valuni Kipos; Fundaci\'o Privada Cellex; CERCA Programme/Generalitat de Catalunya; Spanish Ministry of Economy and Competitiveness, through the 'Severo Ochoa' Programme for Centres of Excellence in R\&D (SEV-2015-0522) and OPTO-SCREEN (TEC2016-75080-R). European Research Council (ERC) project ERIDIAN (713682); European Union project QUIC (Grant Agreement no. 641122); Spanish MINECO projects MAQRO (Ref. FIS2015-68039-P), XPLICA (FIS2014-62181-EXP); Agència de Gestió d’Ajuts Universitaris i de Recerca (AGAUR) project (2014-SGR-1295);

%\bigskip \noindent See \href{link}{Supplement 1} for supporting content.

%Bibliography
%\bibliography{mybib}

\begin{thebibliography}{10}
\newcommand{\enquote}[1]{``#1''}

\bibitem{Cai2007}
X.~Cai and X.~Wang, \enquote{Stochastic modeling and simulation of gene
  networks-a review of the state-of-the-art research on stochastic
  simulations,} IEEE Signal Processing Magazine \textbf{24}, 27--36 (2007).

\bibitem{Click2011}
T.~H. Click, A.~Liu, and G.~A. Kaminski, \enquote{Quality of random number
  generators significantly affects results of monte carlo simulations for
  organic and biological systems,} Journal of Computational Chemistry
  \textbf{32}, 513--524 (2011).

\bibitem{Hall1997}
C.~Hall and B.~Schneier, \enquote{Remote electronic gambling,} in
  \enquote{Computer Security Applications Conference, 1997. Proceedings., 13th
  Annual,}  (IEEE, 1997), pp. 232--238.

\bibitem{Shannon1949}
C.~E. Shannon, \enquote{Communication theory of secrecy systems,} Bell System
  Technical Journal \textbf{28}, 656--715 (1949).

\bibitem{Tajima2007}
A.~Tajima, A.~Tanaka, W.~Maeda, S.~Takahashi, and A.~Tomita, \enquote{Practical
  quantum cryptosystem for metro area applications,} IEEE Journal of Selected
  Topics in Quantum Electronics \textbf{13}, 1031--1038 (2007).

\bibitem{Abellan2015}
C.~Abell\'an, W.~Amaya, D.~Mitrani, V.~Pruneri, and M.~W. Mitchell,
  \enquote{Generation of fresh and pure random numbers for loophole-free bell
  tests,} Phys. Rev. Lett. \textbf{115}, 250403 (2015).

\bibitem{Hensen2015}
B.~Hensen, H.~Bernien, A.~Dr{\'e}au, A.~Reiserer, N.~Kalb, M.~Blok,
  J.~Ruitenberg, R.~Vermeulen, R.~Schouten, C.~Abell{\'a}n, W.~Amaya, 
V.~Pruneri, M.~W.~Mitchell, M.~Markham, D.~J.~Twitchen, D.~Elkouss, S.~Wehner, 
T.~H.~Taminiau, R.~Hanson,  \enquote{Loophole-free bell inequality violation using electron spins
  separated by 1.3 kilometres,} Nature \textbf{526}, 682--686 (2015).

\bibitem{Giustina2015}
M.~Giustina, M.~A. Versteegh, S.~Wengerowsky, J.~Handsteiner, A.~Hochrainer,
  K.~Phelan, F.~Steinlechner, J.~Kofler, J.-{\AA}. Larsson, C.~Abell{\'a}n, W.~Amaya,
 V.~Pruneri, M.~W.~Mitchell, J.~Beyer, T.~Gerrits, A.~E.~Lita, L.~K.~Shalm, 
S.~W.~Nam, T.~Scheidl, R.~Ursin, B.~Wittman, A.~Zeilinger,
\enquote{Significant-loophole-free test of bell?s theorem with
  entangled photons,} Physical Review Letters \textbf{115}, 250401 (2015).

\bibitem{Shalm2015}
L.~K. Shalm, E.~Meyer-Scott, B.~G. Christensen, P.~Bierhorst, M.~A. Wayne,
  M.~J. Stevens, T.~Gerrits, S.~Glancy, D.~R. Hamel, M.~S. Allman, K.~J.~Coakley, 
S.~D.~Dyer, C.~Hodge, A.~E.~Lita, V.~B.~Verma, C.~Lambrocco, E.~Tortorici, A.~L.~Migdall,
 Y.~Zhang, D.~R.~Kumor, W.~H.~Farr, F.~Marsili, M.~D.~Shaw, J.~A.~Stern, C.~Abell{\'a}n, 
W.~Amaya, V.~Pruneri, T.~Jennewein, M.~W.~Mitchell, P.~G.~Kwiat, J.~C.~Bienfang, 
R.~P.~Mirin, E.~Knill, S.~W.~Nam,
\enquote{Strong loophole-free test of local realism,} Physical
  Review Letters \textbf{115}, 250402 (2015).

\bibitem{VonNeumann1951}
J.~Von~Neumann, \enquote{Various techniques used in connection with random
  digits,} Appl. Math Ser \textbf{12}, 36--38 (1951).

\bibitem{Rarity1994}
J.~Rarity, P.~Owens, and P.~Tapster, \enquote{Quantum random-number generation
  and key sharing,} Journal of Modern Optics \textbf{41}, 2435--2444 (1994).

\bibitem{Jennewein2000}
T.~Jennewein, U.~Achleitner, G.~Weihs, H.~Weinfurter, and A.~Zeilinger,
  \enquote{A fast and compact quantum random number generator,} Review of
  Scientific Instruments \textbf{71}, 1675--1680 (2000).

\bibitem{Wahl2011}
M.~Wahl, M.~Leifgen, M.~Berlin, T.~R??hlicke, H.-J. Rahn, and O.~Benson,
  \enquote{An ultrafast quantum random number generator with provably bounded
  output bias based on photon arrival time measurements,} Applied Physics
  Letters \textbf{98}, 171105 (2011).

\bibitem{Gabriel2010}
C.~Gabriel, C.~Wittmann, D.~Sych, R.~Dong, W.~Mauerer, U.~L. Andersen,
  C.~Marquardt, and G.~Leuchs, \enquote{A generator for unique quantum random
  numbers based on vacuum states,} Nature Photonics \textbf{4}, 711--715
  (2010).

\bibitem{Argyris2010}
A.~Argyris, S.~Deligiannidis, E.~Pikasis, A.~Bogris, and
  D.~Syvridis, \enquote{Implementation of 140 Gb/s true random bit generator based on
 a chaotic photonic integrated circuit,} Opt. Express \textbf{18}, 18763 (2010).

\bibitem{Ugajin2017}
K.~Ugajin, Y.~Terashima, K.~Iwakawa, A.~Uchida, T.~Harayama, K.~Yoshimura, and
  M.~Inubushi, \enquote{Real-time fast physical random number generator with a photonic 
integrated circuit,} Opt. Express \textbf{25}, 6511 (2017).

\bibitem{Qi2010}
B.~Qi, Y.-M. Chi, H.-K. Lo, and L.~Qian, \enquote{High-speed quantum random
  number generation by measuring phase noise of a single-mode laser,} Optics
  letters \textbf{35}, 312--314 (2010).

\bibitem{Jofre2011}
M.~Jofre, M.~Curty, F.~Steinlechner, G.~Anzolin, J.~P. Torres, M.~W. Mitchell,
  and V.~Pruneri, \enquote{True random numbers from amplified quantum vacuum,}
  Opt. Express \textbf{19}, 20665--20672 (2011).

\bibitem{Abellan2014}
C.~Abell\'{a}n, W.~Amaya, M.~Jofre, M.~Curty, A.~Ac\'{i}n, J.~Capmany,
  V.~Pruneri, and M.~W. Mitchell, \enquote{Ultra-fast quantum randomness
  generation by accelerated phase diffusion in a pulsed laser diode,} Opt.
  Express \textbf{22}, 1645--1654 (2014).

\bibitem{Yuan2014}
Z.~Yuan, M.~Lucamarini, J.~Dynes, B.~Fr{\"o}hlich, A.~Plews, and A.~Shields,
  \enquote{Robust random number generation using steady-state emission of
  gain-switched laser diodes,} Applied Physics Letters \textbf{104}, 261112
  (2014).

\bibitem{Nie2015}
Y.-Q. Nie, L.~Huang, Y.~Liu, F.~Payne, J.~Zhang, and J.-W. Pan, \enquote{The
  generation of 68 gbps quantum random number by measuring laser phase
  fluctuations,} Review of Scientific Instruments \textbf{86}, 063105 (2015).

\bibitem{Khanmohammadi2015}
A.~Khanmohammadi, R.~Enne, M.~Hofbauer, and H.~Zimmermanna, \enquote{A
  monolithic silicon quantum random number generator based on measurement of
  photon detection time,} IEEE Photonics Journal \textbf{7}, 1--13 (2015).

\bibitem{Abellan2016}
C.~Abellan, W.~Amaya, D.~Domenech, P.~Mu{\~n}oz, J.~Capmany, S.~Longhi, M.~W.
  Mitchell, and V.~Pruneri, \enquote{Quantum entropy source on an inp photonic
  integrated circuit for random number generation,} Optica \textbf{3}, 989--994
  (2016).

\bibitem{Camacho2012}
R.~E. Camacho-Aguilera, Y.~Cai, N.~Patel, J.~T. Bessette, M.~Romagnoli, L.~C.
  Kimerling, and J.~Michel, \enquote{An electrically pumped germanium laser,}
  Opt. Express \textbf{20}, 11316--11320 (2012).

\bibitem{Bie2017}
Y.-Q. Bie, G.~Grosso, M.~Heuck, M.~M. Furchi, Y.~Cao, J.~Zheng, D.~Bunandar,
  E.~Navarro-Moratalla, L.~Zhou, D.~K. Efetov \emph{et~al.}, \enquote{A
  mote2-based light-emitting diode and photodetector for silicon photonic
  integrated circuits,} Nature Nanotechnology  (2017).

\bibitem{Nisan1999}
N.~Nisan and A.~Ta-Shma, \enquote{Extracting randomness: A survey and new
  constructions,} Journal of Computer and System Sciences \textbf{58}, 148 --
  173 (1999).

\end{thebibliography}

% Full bibliography added automatically for Optics Letters submissions
% Note that this extra page will not count against page length
\ifthenelse{\equal{\journalref}{ol}}{%
\clearpage
\bibliographyfullrefs{mybib}
}{}

\end{document}